\documentclass[11pt,a4paper]{amsart}
\usepackage{amsfonts}
\usepackage{hyperref}

\def\={\ =\ }

\theoremstyle{plain}

\numberwithin{equation}{section}

\begin{document}
\title[Chern-Simons matrix models]{Chern-Simons matrix models,
two-dimensional Yang-Mills theory and the Sutherland model}
\date{ \ March 2010 \hfill HWM--10--5 \ , \ EMPG--10--05 \ }
\author{Richard J. Szabo}
\address{\flushleft Department of Mathematics, Heriot-Watt University, Colin
Maclaurin Building, Riccarton, Edinburgh EH14 4AS, UK, and Maxwell Institute
for Mathematical Sciences, Edinburgh, UK}
\email{R.J.Szabo@ma.hw.ac.uk}
\urladdr{}
\thanks{}
\author{Miguel Tierz}
\address{\flushleft Universitat Polit\'{e}cnica de Catalunya, Department de F%
\'{\i}sica i Enginyeria Nuclear, Comte Urgell 187, E-08036 Barcelona, Spain}
\email{csdt.rmt@gmail.com , tierz@brandeis.edu}
\urladdr{}
\curraddr{ }

\begin{abstract}
We derive some new relationships between matrix models of Chern-Simons gauge
theory and of two-dimensional Yang-Mills theory. We show that $q$%
-integration of the Stieltjes-Wigert matrix model is the discrete matrix
model that describes $q$-deformed Yang-Mills theory on $S^{2} $. We
demonstrate that the semiclassical limit of the Chern-Simons matrix model is
equivalent to the Gross-Witten model in the weak coupling phase. We study
the strong coupling limit of the unitary Chern-Simons matrix model and show
that it too induces the Gross-Witten model, but as a first order deformation
of Dyson's circular ensemble. We show that the Sutherland model is
intimately related to Chern-Simons gauge theory on $S^{3}$, and hence to $q$%
-deformed Yang-Mills theory on $S^{2}$. In particular, the ground state
wavefunction of the Sutherland model in its classical equilibrium
configuration describes the Chern-Simons free energy. The correspondence is
extended to Wilson line observables and to arbitrary simply-laced gauge
groups.
\end{abstract}

\maketitle

\section{Introduction and summary of results}

Matrix models have been a subject of much interest in gauge theory for over
three decades~\cite{Brezin:1977sv}. In this paper we study random matrix
models~\cite{Mehta} related to Chern-Simons theory~\cite{cs}, and both
ordinary two-dimensional Yang-Mills theory~\cite{cordesmoore} and its $q$%
-deformed counterpart~\cite{Klimcik:1999kg,Aganagic:2004js}. We will develop
some new relationships between the Chern-Simons matrix models and some of
the matrix models that appear in two-dimensional Yang-Mills theory.

Let us begin by giving the definition of the joint probability distribution $%
P( M) $ for the matrix elements of an $N\times N$ matrix $M$~\cite{Mehta},%
\begin{equation}
P(M)=Z^{-1}_{N}\,\exp \big(-\mathrm{Tr}\,V(M)\big) ~,  \label{matrix}
\end{equation}%
with any potential $V( M)$ such that the partition function $Z_{N}$ exists.
The integration of $\left( \ref{matrix}\right) $ over parameters related to
the eigenvectors of $M$ leads to the well-known joint probability
distribution of the eigenvalues~\cite{Mehta}%
\begin{equation}
P(x_{1},\dots ,x_{N})=Z_{N}^{-1}\,\prod_{i<j}\, \left\vert
x_{i}-x_{j}\right\vert ^{\beta }~ \prod_{i=1}^{N}\, \exp\big(-V(x_{i})\big) %
~.  \label{rmt}
\end{equation}%
The level repulsion in $\left( \ref{rmt}\right) $, described by the
Vandermonde determinant, originates as the Jacobian of the transformation
when passing from integration over independent matrix elements to
integration over the smaller space of $N$ eigenvalues. The integer $\beta
=1,2,4$ describes the symmetry of the ensemble (orthogonal, unitary and
symplectic, respectively).

These symmetries have been extended using the Cartan classification of
symmetric spaces~\cite{Altland:1997zz} (see also~\cite%
{Caselle:2003qa,Zirnbauer,Zirnbauer:2010gg}). For example, the possible
Jacobians of the transformations to radial coordinates are given by~\cite%
{Caselle:2003qa}%
\begin{eqnarray}
J^{\left( 0\right) }(x) &=&\prod\limits_{\alpha \in \Delta ^{+}}\,\left(
x_{\alpha }\right) ^{m_{\alpha }}\ , \\[4pt]
J^{\left( -\right) }(x) &=&\prod\limits_{\alpha \in \Delta ^{+}}\,\left(
a^{-1}\,\sinh (x_{\alpha })\right) ^{m_{\alpha }}\ ,  \notag \\[4pt]
J^{\left( +\right) }(x) &=&\prod\limits_{\alpha \in \Delta ^{+}\,}\left(
a^{-1}\,\sin (x_{\alpha })\right) ^{m_{\alpha }}\ ,  \notag
\end{eqnarray}%
for the various types of symmetric spaces with zero, negative and positive
constant curvature, respectively, and for an arbitrary non-zero constant $a$%
. The products are taken over all positive roots of the restricted root
lattice, with $m_{\alpha }$ the multiplicity of the root vector $\alpha \in
\Delta ^{+}$, and $x_{\alpha }:=(x,\alpha )=\sum_{i}\,x_{i}\,\alpha ^{i}$
are canonical coordinates on a maximal abelian subalgebra of the tangent
space.

The case of the $A_{N-1}$ root system and zero curvature leads to the
well-known Gaussian matrix model distribution (we write only the Hermitian
case)%
\begin{equation}
P^{(0)}(x_{1},\dots ,x_{N})=Z_{N}^{-1}\,\prod_{i<j}\,\left(
x_{i}-x_{j}\right) ^{2}~\prod_{i=1}^{N}\,\exp \big(-x_{i}^{2}/2\big)~,
\label{Gaussian}
\end{equation}%
whereas the same root lattice but in the case of negative curvature yields%
\begin{equation}
P^{(-)}(x_{1},\dots ,x_{N})=Z_{N}^{-1}\,\prod_{i<j}\,\sinh ^{2}\big(%
a\,(x_{i}-x_{j})\big)~\prod_{i=1}^{N}\,\exp \big(-x_{i}^{2}/2\big)~.
\label{Chern-Simons}
\end{equation}%
In the B-model topological string theory, this is the matrix model that
describes $U(N)$ Chern-Simons gauge theory on $S^{3}$~\cite%
{Marino:2002fk,Aganagic:2002wv} if $a=\sqrt{g_{s}}$, where $g_{s}$ is the
string coupling constant which is related to the usual integer Chern-Simons
level $k$ by 
\begin{equation}
g_{s}=\frac{2\pi {\,\mathrm{i}\,}}{k+N}\ .  \label{gskN}
\end{equation}%
The exponential mapping $x_{i}=\log u_{i}/a$ brings $\left( \ref%
{Chern-Simons}\right) $ into the usual form $\left( \ref{rmt}\right) $ (with 
$\beta =2$), and hence one can apply the orthogonal polynomial method of
random matrix theory~\cite{Mehta} to solve the model \cite{Forr,Tierz}. The
orthogonal polynomials here are the Stieltjes-Wigert polynomials~\cite%
{Forr,Tierz,DTierz} or, if we work with a unitary matrix model~\cite%
{Okuda:2004mb}, the Rogers-Szeg\H{o} polynomials~\cite{DTierz}.

Chern-Simons theory on $S^3$ is equivalent to $q$-deformed Yang-Mills theory
on $S^2$, with the identification $q:={\,\mathrm{e}}\,^{-g_s}$. This result
can be derived at weak coupling either directly via localization of the
three-dimensional gauge theory by regarding $S^3$ as a Seifert manifold
through the Hopf fibration $S^3\to S^2$~\cite{Beasley:2005vf,Blau:2006gh},
or by recasting the two-dimensional gauge theory as a sum over instantons
and explicitly demonstrating its equivalence with the Chern-Simons matrix
model~\cite{Caporaso:2005ta,Griguolo:2006kp}. At weak coupling, the
two-dimensional $U(N)$ gauge theory thus reproduces the perturbative A-model
topological string partition function $Z_{\mathrm{top}}$ for the resolved
conifold geometry in the large $N$ limit. The Stieltjes-Wigert polynomial is
the average of the characteristic polynomial in the matrix model, and hence
describes B-brane amplitudes on the conifold~\cite{Okuyama:2006eb}.

In the next section we show that $q$-integration of the Stieltjes-Wigert
matrix model directly gives the discrete matrix model that describes the
strong coupling expansion of $q$-deformed two-dimensional Yang-Mills theory.
In the large $N$ limit, the strong coupling series has zero radius of
convergence~\cite{Caporaso:2005ta}. This is in sharp contrast to ordinary
(undeformed) Yang-Mills theory on $S^2$, which undergoes a third order phase
transition at large $N$~\cite{Douglas:1993iia} and possesses a double
scaling limit which lies in the universality class of the Gross-Witten
unitary matrix model~\cite{Gross:1994mr}. The double scaling limit of the
Gross-Witten model is also of interest in the study of unitary matrix models
of string theory in zero dimensions~\cite{Periwal}, of the solution of $%
SU(2) $ Seiberg-Witten theory~\cite{Dijkgraaf:2002vw}, and of Type~0A and 0B
string theories in one dimension~\cite{Kms} where it is argued to describe
the universality class of pure two-dimensional supergravity.

In the next section we will demonstrate that both the weak and strong
coupling limits of the finite $N$ Chern-Simons matrix models are also
governed by the Gross-Witten model~\cite{GW}. Recall that this is the
unitary one-matrix model which arises as the one-plaquette reduction of the
combinatorial quantization of Yang-Mills theory in \emph{infinite}
spacetime. In two dimensions the reduction is exact and described by the
partition function~\cite{GW}%
\begin{align}
Z_N^{\mathrm{GW}}(\alpha) & := \int_{U(N)}\, \mathrm{d}U~{\exp }\left(
-\alpha \,\mathrm{Tr}\big( U+U^{\dag }\big) \right)  \label{GW} \\[4pt]
& =\int_0^{2\pi}~ \prod\limits_{i=1}^{N} \, \mathrm{d}\theta _{i}~ \mathrm{e}%
^{-2\alpha\, \cos \theta_i }~\prod\limits_{i<j}\, \sin ^{2}\left( \frac{%
\theta _{i}-\theta _{j}}{2}\right) \ ,  \notag
\end{align}%
where $\mathrm{d} U$ denotes the bi-invariant Haar measure for integration
over the unitary group $U(N)$. This matrix model belongs to the class of
symmetric spaces associated with the $A_{N-1}$ root system and positive
curvature.

We will thus show that the $q\rightarrow 1$ limit of the unitary
Chern-Simons matrix models are equivalent to two-dimensional Yang-Mills
theory on the \emph{plane} $\mathbb{R}^2$. Identifying $\alpha=\frac1{g_s}=%
\frac{k+N}{2\pi{\,\mathrm{i}\,}}$ and using the exact integration of (\ref%
{GW}) in the weak-coupling phase~\cite[eq.~(5.25)]{Semenoff:1996vm}, we can
write down a simple all-orders expression for the $U(N)$ Chern-Simons
partition function $Z_{N,k}(S^3)$ on $S^3$ in the weak-coupling limit as 
\begin{equation}
\lim_{k\to\infty}\, Z_{N,k}\left(S^3\right) = \left(\frac{2\pi{\,\mathrm{i}\,%
} {\,\mathrm{e}}\,^{2(k+N)/\pi{\,\mathrm{i}\,}}}{N\, (k+N)}
\right)^{N^2/2}\, \left(1+(-1)^{-N^2/2}~{\,\mathrm{e}}\,^{-N\,(k+N)/\pi{\,%
\mathrm{i}\,}} \right)^N \ .
\end{equation}
This exact expression should reproduce the topological string partition
function $Z_{\mathrm{top}}$ in the large $N$ limit to leading orders of
perturbation theory.

As the unitary one-matrix model (\ref{GW}) undergoes a phase transition to a
strong coupling phase in the large $N$ limit, while the $q$-deformed gauge
theory on $S^2$ always remains in its weak coupling phase, the nature of
this relationship must be drastically different at strong coupling. We will
show that the $q\to0$ limit of the unitary Chern-Simons matrix model is
equivalent to Dyson's circular ensemble, together with an infinite tower of
higher Casimir deformations corresponding to multicritical extensions~\cite%
{Periwal}, the lowest order of which is described by the Gross-Witten model.
These deformations are analogous to those which arise in Yang-Mills theory
on the noncommutative torus~\cite{Paniak:2003gn,Paniak:2003xm}. This feature
puts the $q$-deformed gauge theory into the context of noncommutative
deformations of Yang-Mills theory~\cite{Szabo:2001kg}.

In the last section we show that the celebrated Sutherland model~\cite{Suth}%
, a central model in the theory of one-dimensional integrable systems, is
also directly related to Chern-Simons gauge theory. This provides another
connection between two-dimensional Yang-Mills theory and Chern-Simons
theory, and moreover between integrable models and Chern-Simons theory on
the three-sphere $S^{3}$. The Sutherland model is the exactly solvable
system on a circle of circumference $L$ defined by the $N$-particle quantum
Hamiltonian operator~\cite{Suth}%
\begin{equation}  \label{SuthHam}
H=-\sum_{i=1}^{N}\, \frac{\partial ^{2}}{\partial q_{i}^{2}}+\frac{2\lambda
\left( \lambda -1\right) \pi ^{2}}{L^{2}}\, \sum_{i<j}\, \left(\sin\frac{\pi
\left( q_{i}-q_{j}\right) }{L}\right)^{-2} \ ,
\end{equation}%
with the ground state wavefunction 
\begin{equation}
\Psi _{0}( q_{1},\dots,q_{N};\lambda,L) =\prod\limits_{i<j}\, \left( \sin 
\frac{\pi \left( q_{i}-q_{j}\right) }{L}\right) ^{\lambda } \ .  \label{Suth}
\end{equation}

The relationship follows immediately when the model is considered in a
simple fixed crystalline configuration. As we will discuss, this is not an
artificial configuration, and it corresponds to the classical equilibrium
state of the Sutherland model which is naturally reached in the strong
coupling limit $\lambda\to\infty$. We will show that the probability density
distribution, evaluated in an equally spaced configuration, reproduces the
partition function for Chern-Simons gauge theory on $S^{3}$ with gauge group 
$SU(N)$, which is one of the simplest quantum topological invariants that
can be obtained from Chern-Simons theory~\cite{cs}. We demonstrate that this
correspondence extends to Wilson line observables, and hence to other
quantum topological invariants, by considering excited states of the
Sutherland model associated to particular non-equilibrium configurations of
the particles. It also holds between generalized Calogero-Sutherland models
and Chern-Simons theory with other simply-laced gauge groups (associated to
orthogonal and symplectic ensembles).

Recall that the Chern-Simons partition function $Z_{N,k}(S^3)$ (or
equivalently the free energy $F_{N,k}(S^{3})=\log Z_{N,k}(S^{3})$) is
central in the study of topological strings, for example. The matrix model
description is useful in topological string theory, and its physical
applications have been exploited mainly in that context thus far. Although
there have been extensive applications of Chern-Simons theory to condensed
matter physics (most notably to the fractional quantum Hall effect), our
application here deals with a many-body description of Chern-Simons theory
in its non-abelian version, at least for certain simple three-manifolds such
as $S^{3}$. Given the natural appearance of random matrix theory, and its
connection with exactly solvable models in one dimension~\cite{Tierz} and
with Laughlin wavefunctions~\cite{T2}, it seems worthwhile to further
explore this many-body description in itself and its possible role in
condensed mater physics.

This correspondence also gives an intriguing new relationship between
two-dimensional Yang-Mills theory and certain one-dimensional integrable
systems, which is rather different in spirit from previous relationships~%
\cite{Gorsky:1993pe,Minahan:1993mv} that considered wavefunctions of
Calogero-Sutherland and Calogero-Moser models as reductions of those for the
undeformed gauge theory on a cylinder with appropriate Wilson line
insertions. The $q$-deformed gauge theories on $S^2$ may thus provide
computationally useful means for exploring various aspects of
one-dimensional exactly solvable models. In particular, as the large $N$
deformed gauge theory on $S^2$ exhibits no phase transitions, there is a
natural and simple well-defined large $N$ field theory limit of the
Sutherland system which is equivalent to a generalized Calogero-Sutherland
model based on the $su(\infty)$ Lie algebra. In~\cite{Gorsky:1993pe} this
limit is instead realized by replacing $SU(\infty)$ with the centrally
extended $SU(N)$ loop group, which is related to Yang-Mills theory on an
infinite cylinder with Wilson line insertions. It would be interesting to
better understand the explicit relationship between the two approaches,
given the known equivalence between the partition functions of $q$-deformed
Yang-Mills theory on $S^2$ and of ordinary Yang-Mills theory on a cylinder
with trivial holonomies around the two boundary circles of the cylinder~\cite%
{dH}.

\subsection*{Acknowledgments}

This work was supported in part by grant ST/G000514/1 \textquotedblleft
String Theory Scotland\textquotedblright\ from the UK Science and Technology
Facilities Council. MT thanks Mark Adler for hospitality at the Mathematics
Department at Brandeis University, and the Department of Mathematics at
Heriot-Watt University for a productive stay.

\section{From Chern-Simons theory to the Gross-Witten model}

\subsection{Another derivation of ${q}$-deformed Yang-Mills theory}

Let us begin by computing the $q$-integration of the Stieltjes-Wigert matrix
model. We will use Jackson's integral, which in the single variable case is
given by~\cite{AAR,K}%
\begin{equation}
\int_{0}^{\infty }\,\mathrm{d}_{q}u~ w(u) =(1-q)\, \sum_{n=-\infty }^{\infty
}\, w(q^{n})\, q^{n} \ .
\end{equation}%
If the function $w(u)$ is the weight function of the Stieltjes-Wigert matrix
model, i.e. the log-normal distribution $w(u) =\exp \left( -\log
^{2}u/2g_{s}\right) $, then the $q$-integration gives%
\begin{equation}
\int_{0}^{\infty }\,\mathrm{d}_{q}u ~ \mathrm{e}^{-{\log ^{2}u}/{2g_{s}}%
}=\left( 1-q\right) \, \sum_{n=-\infty }^{\infty }\, q^{n^{2}/2+n} \ ,
\label{qtheta}
\end{equation}%
with the usual identification $q={\,\mathrm{e}}\,^{-g_{s}}$. The right-hand
side of $\left( \ref{qtheta}\right) $ is proportional to the theta-function 
\begin{equation}
\Theta _{00}(z|q)=\sum_{n=-\infty}^\infty\, q^{n^2/2}\,z^n  \label{Theta}
\end{equation}
at the particular value $z=q.$ This theta-function is also the weight
function of the unitary matrix model that describes Chern-Simons gauge
theory~\cite{Okuda:2004mb,DTierz}. We shall see below how this feature
relates the Chern-Simons matrix model with the Gross-Witten model.

Let us now compute the $q$-integration of the log-normal weight function in
the multivariable case. For the Stieltjes-Wigert matrix model, we then have%
\begin{eqnarray}
&& \int_{0}^{\infty }~ \prod\limits_{i=1}^{N}\, \frac{\mathrm{d}_{q}u_{i}}{%
2\pi }~{\,\mathrm{e}}\,^{-{\log ^{2}u_{i}}/{2g_{s}}}~ \prod\limits_{i<j}\,
\left( u_{i}-u_{j}\right) ^{2} \\
&& \hspace{4cm} \= \left(\frac{1-q}{2\pi} \right)^{N}\,
\sum_{n_{1},\dots,n_{N}=-\infty }^{\infty }\, q^{\sum_{i}\, (
n_{i}^{2}/2+n_{i})}~ \prod\limits_{i<j}\, \left( q^{n_{i}}-q^{n_{j}}\right)
^{2} \ ,  \notag
\end{eqnarray}%
which coincides with the partition function of the $q$-deformed
two-dimensional Yang-Mills model on $S^{2}$ with gauge group $U(N)$~\cite%
{Klimcik:1999kg,Aganagic:2004js}. Recall that, using solutions of the
log-normal moment problem, the equivalence between the continuum and
discrete matrix models was demonstrated in this case in~\cite{mtierz}, and
it reads 
\begin{eqnarray}
&&\left( \frac{g_{s}}{2\pi }\right) ^{-{N}/{2}}\, \int_{-\infty }^{\infty }~
\prod\limits_{i=1}^N\, \frac{\mathrm{d} x_{i}}{2\pi }~ {\,\mathrm{e}}\,^{-{%
x_{i}^{2}}/{2g_{s}}}~\prod\limits_{i<j}\, \left( 2\sinh \Big( \frac{%
x_{i}-x_{j}}{2}\Big) \right) ^{2}  \label{sinhdisc2} \\
&& \hspace{3cm} \= A_{N}(q)\, \sum_{n_{1},\dots,n_{N}=-\infty }^{\infty }\, {%
\,\mathrm{e}}\,^{-\frac{g_{s}}{2}\, \sum_{i}\,
n_{i}^{2}}~\prod\limits_{i<j}\, \bigg( 2\sinh \Big( \frac{g_{s}}{2}\,(
n_{i}-n_{j}) \Big) \bigg) ^{2}  \notag
\end{eqnarray}%
where%
\begin{equation}
A_{N}(q):=\left(\, {\frac{q^{-(1-2N+3N^{2})/2}}{\big(-q^{3/2-N}\big)_{\infty
}\, \big(-q^{N-1/2}\big)_{\infty }\, (q)_{\infty }}}\,\right) ^{N} \ ,
\end{equation}%
with $(q)_0:=1$, $(q)_m:=\prod_{1\leq j\leq m}\, \big(1-q^j\big)$ for $m\in%
\mathbb{N}$, and $(q)_\infty:=\prod_{j\in\mathbb{N}}\, \big(1-q^j\big)$.
Hence $q$-integration and ordinary integration of the Stieltjes-Wigert
distribution gives the same result (up to overall normalization), the $U(N)$
Chern-Simons partition function on $S^{3}$.

\subsection{Weak-coupling limit}

We will now demonstrate that the Chern-Simons matrix model yields, in the
weak-coupling limit $g_{s}\rightarrow 0$ $\left( q=\mathrm{e}%
^{-g_{s}}\rightarrow 1\right) $, the Gross-Witten model (\ref{GW}) with
coupling $g=\alpha ^{-1}\rightarrow 0.$ For this, we note that the
distinctive log-square behaviour of the weight function $w(u) =\exp \left(
-\log ^{2}u/2g_{s}\right) $ for the Stieltjes-Wigert matrix model also
constitutes the natural definition of $q$-growth~\cite{R,Gelfand}. We use
results of~\cite{R} to study its relationship with the different definitions
of $q$-exponential functions, and in turn to better understand its $%
q\rightarrow 1$ limit. The crux of our analysis is that the weight function
of the Stieltes-Wigert matrix model can be given in terms of $q$-exponential
functions, but without the usual rescaling $u\rightarrow \left( 1-q\right) u$
that is necessary to recover ordinary calculus from $q$-deformed calculus
when $q\rightarrow 1$~\cite{AAR,K}.

As we mentioned in the previous section, the matrix models that appear in
Chern-Simons theory can be solved exactly by using orthogonal polynomials.
These orthogonal polynomials are completely characterized by a discrete
scaling symmetry satisfied by the weight function~\cite{mtierz}%
\begin{equation}
w(q\, u)=\sqrt{q}\,u\,w(u) \ .  \label{func}
\end{equation}%
This self-similarity property is called the $q$-Pearson equation, and it
uniquely determines the orthogonal polynomials up to normalization~\cite%
{mtierz}.

Let us begin by briefly reviewing $q$-deformed expressions for exponential
functions and how they tend to their classical counterparts in the limit $%
q\rightarrow 1$~\cite{AAR,K}. For $u\in\mathbb{C}$, the two main definitions
of a $q$-exponential function in the literature are given by 
\begin{align}
e_{q}(u)& =\sum_{j=0}^{\infty }\,\frac{u^{j}}{\left(q\right)_{j}}=\frac{1}{%
\left(u;q\right) _{\infty }} \ , \\[4pt]
E_{q}(u)& =\sum_{j=0}^{\infty }\, \frac{q^{j\,(j-1)/2}\, u^{j}}{\left(
q\right)_{j}}=\left( -u;q\right) _{\infty }=\big( e_{q}(-u)\big) ^{-1} \ , 
\notag
\end{align}%
where $(u;q)_\infty:=\prod_{j\in\mathbb{N}_0}\, \big(1-u\,q^j\big)$. The
connection with the usual exponential function is given by the functional
equation 
\begin{equation}
e_{q}(u)\, E_{q}(-u)=1
\end{equation}%
and the weak-coupling limits%
\begin{equation}
\lim_{q\rightarrow 1}\, e_{q}\big(\left( 1-q\right) u\big)%
=\lim_{q\rightarrow 1}\, E_{q}\big(\left( 1-q\right) u\big)= {\,\mathrm{e}}%
\,^u \ .  \label{limit}
\end{equation}

We now recall that the functional equation $\left( \ref{func}\right) $
solves the matrix model~\cite{mtierz}. Furthermore, one can show~\cite{R}
that the entire function $f(u)=E_{q}(u)$ satisfies $(1+u)\, f(q\, u)=f(u)$
and the entire function $g(u)=e_{q^{-1}}(-u^{-1})$ satisfies $%
u\,g(u)=(1+u)\, g(u)$. It follows that the entire function%
\begin{equation}
h(u)=f(u)\, g(u)=\left( -u;q\right) _{\infty }\, \left( -q\, u^{-1};q\right)
_{\infty }
\end{equation}%
is a solution of the $q$-difference equation $u\, h(q\, u)=h(u)$. By
replacing $q\rightarrow q^{-1}$ and rescaling $u\rightarrow q\, u$, the
functional equation for $h(u)$ reads%
\begin{equation}
h(q\,u)=q\,u\, h(u) \ .  \label{func2}
\end{equation}%
This is essentially the $q$-Pearson equation $\left( \ref{func}\right)$, but
with $q$ instead of $\sqrt{q}$ appearing the right-hand side, and hence
numerical prefactors will be different if we use (\ref{func2}) instead of (%
\ref{func}) as the defining characterization of the Stieltjes-Wigert weight
function $w(u)$.

Thus the explicit $q$-deformed expression that satisfies $\left( \ref{func2}%
\right) $ is given by%
\begin{equation}
w(u)=E_{q^{-1}}(q\,u)\,e_{q}\big(-(q\,u)^{-1}\big)\ .
\end{equation}%
An elementary manipulation shows $E_{q^{-1}}(u)=e_{q}(u)$, and so from (\ref%
{limit}) the weak-coupling limit $q\rightarrow 1$ is given by%
\begin{equation}
\lim_{q\rightarrow 1}\,w(u)=\exp \left( -\frac{u}{1-q}\right) \,\exp \left( -%
\frac{u^{-1}}{1-q}\right) =\exp \left( -\frac{1}{1-q}\,\left(
u+u^{-1}\right) \right) \ .
\end{equation}%
In the unitary Chern-Simons matrix model, we substitute $u=\mathrm{e}^{{\,%
\mathrm{i}\,}\theta }$ with $0\leq \theta \leq 2\pi $ to get%
\begin{equation}
\lim_{q\rightarrow 1}\,w\big(u={\,\mathrm{e}}\,^{{\,\mathrm{i}\,}\theta }%
\big)=\exp \left( -\frac{2}{1-q}\,\cos \theta \right) \ ,  \label{limitq1}
\end{equation}%
and we arrive at the Gross-Witten model $\left( \ref{GW}\right) $ with $%
\alpha \propto \frac{1}{1-q}$. Since $q\rightarrow 1$ ($g_{s}\rightarrow 0$)
we can write $\alpha \simeq \frac{1}{g_{s}}$. The factor of $2$ in $\left( %
\ref{limitq1}\right) $ is not accurate since we have used $\left( \ref{func2}%
\right) $ instead of $\left( \ref{func}\right) $.

The same result is contained in the $q\rightarrow 1$ limit of the $q$%
-Hermite polynomials studied by Ismail, Stanton and Viennot in \cite{ISV}.
They consider polynomials with the same orthogonality properties as the ones
that solve the Chern-Simons matrix model (see the discussion of the $%
q\rightarrow 0$ limit below), but then apply a rescaling of the variable $%
x\rightarrow \left( \sqrt{1-q}/2\right) x$ in order to obtain the correct $%
q\to1$ limit in their weight function $\upsilon(x,q)$ as mentioned above. If
we undo this rescaling and take into account that with their definitions $%
x=\cos( \theta /2) $, then their result%
\begin{equation}
\lim_{q\rightarrow 1}\, \upsilon( x,q) =\exp \left( -x^{2}/2\right)
\end{equation}%
leads in our case to%
\begin{equation}
\lim_{q\rightarrow 1}\, \upsilon(x,q)= \exp \left( -\frac{2}{1-q}\, \cos ^{2}%
\frac{\theta }{2}\right) = {\,\mathrm{e}}\,^{-1/(1-q)}\, \exp \left( -\frac{%
\cos \theta }{1-q}\right) \ .
\end{equation}

\subsection{Strong-coupling limit}

Let us now study the connection between the Chern-Simons matrix model and
the Gross-Witten model in the strong-coupling limit $g_s\to\infty$ ($q={\,%
\mathrm{e}}\,^{-g_s}\rightarrow 0$). Recall first that both the
Stieltjes-Wigert polynomials, defined on the positive real half-line, and
the Rogers-Szeg\H{o} polynomials, defined on the unit circle~\cite{Szego},
solve the matrix models that appear in Chern-Simons theory~\cite%
{Tierz,DTierz}. In the Hermitian case this was demonstrated in~\cite{Tierz},
while the unitary matrix model was proposed in~\cite{Okuda:2004mb} (but
without consideration of the associated Rogers-Szeg\H{o} polynomials). The
explicit connection between the two models was demonstrated in~\cite{DTierz}%
. The direct relationship between the two systems of orthogonal polynomials
is well-known~\cite{AN}. Therefore, the partition function for Chern-Simons
gauge theory on $S^{3}$ with gauge group $U(N)$ can be represented both as
the Hermitian matrix model%
\begin{equation}
Z_{N,k}\left( S^{3}\right) =\int\, [\mathrm{d} M]~\exp\left(-{\frac{1}{2g_{s}%
}}\, {\mathrm{Tr}}\, (\log M)^{2}\right)~,  \label{SWMM}
\end{equation}%
with $[\mathrm{d} M]$ the natural invariant measure for integration over the
space of $N\times N$ Hermitian matrices, and as the unitary matrix model~%
\cite{Okuda:2004mb}%
\begin{equation}
Z_{N,k}\left( S^{3}\right) = \int_{U(N)}\, \mathrm{d}U~\det {\Theta
_{00}(U|q)} \ ,
\end{equation}%
where the theta-functions $\Theta _{00}({\,\mathrm{e}}\,^{{\,\mathrm{i}\,}%
\theta_i }|q)$ are defined on eigenvalues of unitary matrices $U$ by (\ref%
{Theta}).

The Rogers-Szeg\H{o} polynomials are defined by \cite{Szego}%
\begin{equation}
H_{n}(z|q):=\sum_{j=0}^{n}\,\left[ 
\begin{matrix}
{n} \\ 
{j}%
\end{matrix}%
\right] _{q}\,z^{j}\ ,\qquad \left[ 
\begin{matrix}
{n} \\ 
{j}%
\end{matrix}%
\right] _{q}:=\frac{(q)_{n}}{(q)_{j}\,(q)_{n-j}}\ .
\end{equation}%
They satisfy an orthogonality relation on the unit circle given by%
\begin{equation}
\oint_{|z|=1}~{\frac{\mathrm{d}z}{2\pi {\,\mathrm{i}\,}z}}~H_{m}\big(%
-q^{-1/2}\,\overline{z}\ \big|\ q\big)\,H_{n}\big(-q^{-1/2}\,z\ \big|\ q\big)%
\,\Theta _{00}(z|q)={\frac{(q)_{m}}{q^{m}}}\,\delta _{m,n}\ .  \label{ort}
\end{equation}%
The orthogonality coefficients $h_{m}=\frac{(q)_{m}}{q^{m}}$ are those of
the Stieltjes-Wigert polynomials, that lead to the Chern-Simons partition
function~\cite{Tierz}.\footnote{%
These are the same orthogonality coefficients that were considered in~\cite%
{DTierz}. The $q^{-m}$ term differs from the coefficients of~\cite{Tierz},
but it only contributes a phase to the partition function. See~\cite{Tierz}
and the discussion of framing in Section~\ref{Framing} below for more
details.}

The weight function appearing in $\left( \ref{ort}\right) $ is $\omega
(z)=\Theta _{00}(z|q)$ which can be expanded in powers of $q$ and $z$ using (%
\ref{Theta}). Since here $z={\,\mathrm{e}}\,^{{\,\mathrm{i}\,}\theta }$
lives on the unit circle, the expansion reads%
\begin{equation}
\omega (\cos \theta )=1+2\,\sqrt{q}\,\cos \theta +2\,\sum_{n\geq
2}\,q^{n^{2}/2}\,\cos n\,\theta \ .  \label{limitq0}
\end{equation}%
The first two terms in (\ref{limitq0}) give the first order approximation of
the weight function for the Chern-Simons matrix model as $q\rightarrow 0$.
This result is very different from the $q\rightarrow 1$ behaviour. The
leading constant term in $\left( \ref{limitq0}\right) $ corresponds to
Dyson's circular ensemble~\cite{Mehta}. The first order correction for small 
$q$ is given by the weight function of the Gross-Witten model (\ref{GW})
with $\alpha =\sqrt{q}={\,\mathrm{e}}\,^{-g_{s}/2}\rightarrow 0$.
Corrections to this term are given by higher powers of $\cos \theta $.

These behaviours are similar to those obtained in~\cite{MI} for a unitary
matrix model with a one-parameter family of weights which defines a $q$%
-deformation of the circular ensemble. The $a=0$ member of this family is
the weight function~\cite[eq.~(3)]{MI}%
\begin{equation}
\upsilon_0( \theta,q) =\left|\big( \sqrt{q}~{\,\mathrm{e}}\,^{{\,\mathrm{i}\,%
}\theta }\, ;\, q\big)_\infty\right|^{2} \ .
\end{equation}%
In the limit $q\rightarrow 0$ it tends to Dyson's circular ensemble and in
the limit $q\rightarrow 1$ to the Gross-Witten model. This weight is very
similar to that given by the theta-function above (although the two infinite
product expansions are different).

\subsection{Comparison with multicritical and noncommutative deformations}

If one keeps only a finite number of terms in the expansion (\ref{limitq0}),
then the corresponding matrix model is the multicritical polynomial
generalization of the Gross-Witten model discussed in~\cite{Periwal}. These
results have been generalized and unitary matrix models of the form%
\begin{equation}
Z_N^{\mathrm{mult}}(t)=\int_{U(N)}\, \mathrm{d} U~ \exp \Big(\,
\sum_{j=-\infty }^{\infty }\, t_{j}\,\mathrm{Tr}\,U^{j}\,\Big)
\end{equation}%
are of interest in various areas of field theory (see~\cite{Morozov:2009jv}
for a recent review). In general, the coupling parameters $t_{j}$ are
restricted by solutions of Virasoro constraints on the partition function $%
Z_N^{\mathrm{mult}}(t)$. However, if one chooses them to be the coefficients
of a $q$-series, and in particular those of the theta-function in (\ref%
{limitq0}), then one obtains the Chern-Simons matrix model.

Our realization of the unitary matrix model describing Chern-Simons theory
as a certain $q$-deformation of the Gross-Witten model is also reminiscent
of what occurs in two-dimensional $U(N)$ Yang-Mills theory on a
noncommutative torus. In~\cite{Paniak:2003gn} it is shown that the
weak-coupling limit of the partition function in this case can be regarded
as coming from a modification of ordinary gauge theory by the addition of
infinitely many higher Casimir operators to the action. The addition of
higher Casimir operators to ordinary, two-dimensional Yang-Mills theory
leads to generalized Yang-Mills theories~\cite{cordesmoore}.

The relationship between these noncommutative deformations and our $q$%
-deformations is most transparent in the combinatorial quantization of the
gauge theory described in~\cite{Paniak:2003xm}, which yields a discrete
family of unitary matrix models parametrized by $N$-th roots of unity $\zeta 
$ and partition functions given by~\cite[eq.~(92)]{Paniak:2003xm}%
\begin{equation}
Z_{N,\zeta }^{\mathrm{NC}}(\alpha )=\sum_{R}\,\frac{1}{\dim R}~\int_{U(N)}\,%
\mathrm{d}U~\chi _{R}(U)\,\exp \left( -\alpha \,\mathrm{Tr}\big(\zeta \,U+%
\overline{\zeta }\,U^{\dag }\big)\right) \ ,  \label{fuzzy}
\end{equation}%
where the sum runs over all irreducible unitary representations $R$ of the
gauge group $U(N)$ with dimension $\dim R$ and characters $\chi _{R}$. The
truncation of the sum in (\ref{fuzzy}) to the trivial representation is
independent of $\zeta $~\cite[\S \S 5.2.1]{Semenoff:1996vm} and corresponds
to the Gross-Witten reduction of the combinatorial quantization of ordinary
gauge theory in two dimensions. The noncommutative gauge theory partition
function thus generalizes that of ordinary Yang-Mills theory by perturbing
it by a sum over non-trivial representations of the unitary group. Again, in
contrast to the $q$-deformed gauge theory on $S^{2}$, this matrix model
admits large $N$ phase transitions and non-trivial double scaling limits
which converge to the continuum (noncommutative) gauge theories~\cite%
{Paniak:2003xm,Szabo:2001kg}.

\section{From the Sutherland model to Chern-Simons theory}

\subsection{Derivation of the Chern-Simons free energy}

Let us now establish the relationship between the Sutherland model and the
Chern-Simons matrix model that we mentioned in the first section. We shall
pay particular attention to the square of the ground state wavefunction (\ref%
{Suth}) for the value $\lambda =\frac{1}{2}$ of the Sutherland coupling,
which is given by%
\begin{equation}
P(q_{1},\dots ,q_{N};L)=\big (\Psi _{0}(q_{1},\dots ,q_{N};\mbox{$\frac12$}%
,L)\big)^{2}=\prod\limits_{i<j}\,\sin \frac{\pi \left( q_{i}-q_{j}\right) }{L%
}\ ,  \label{Suthwave}
\end{equation}%
and evaluate this probability density in the fixed and equally-spaced
configuration $q_{i}=c-i$, with $c\in \mathbb{R}$ an arbitrary constant.
Then we easily find%
\begin{align}
P(c-1,\dots ,c-N;L)& =\prod\limits_{i<j}\,\sin \frac{\pi \left( i-j\right) }{%
L}  \label{evaluation} \\[4pt]
& =\prod\limits_{j=1}^{N-1}\,\left( \sin \frac{\pi }{L}\,\sin \frac{2\pi }{L}%
\,\cdots \,\sin \frac{\pi \left( N-j\right) }{L}\right)
=\prod\limits_{j=1}^{N-1}\,\sin ^{N-j}\left( \frac{\pi }{L}\,j\right) \ . 
\notag
\end{align}

This is proportional to the Chern-Simons partition function on $S^{3}$~\cite%
{cs} once we identify the arbitrary length parameter $L$ as 
\begin{equation}
L=k+N  \label{LkN}
\end{equation}%
in terms of Chern-Simons parameters, the integer level $k$ and the rank of
the gauge group $N.$ However, there is an additional factor $\left(
k+N\right) ^{-N/2}$ in front of the product term in the Chern-Simons
partition function~\cite{cs}. Thus we require a renormalization of the
Sutherland wavefunction $\left( \ref{Suth}\right) $ by multiplying it with
the parameter $L^{-N/4}$. Then one has%
\begin{equation}
\widetilde{\Psi }_{0}(q_{1},\dots ,q_{N};\mbox{$\frac12$},L):=\frac{1}{%
L^{N/4}}\,\prod\limits_{i<j}\,\left( \sin \frac{\pi \left(
q_{i}-q_{j}\right) }{L}\right) ^{1/2}  \label{waveren}
\end{equation}%
with 
\begin{equation}
\big(\,\widetilde{\Psi }_{0}(1,\dots ,N;\mbox{$\frac12$},k+N)\,\big)^{2}={Z}%
_{N,k}\left( S^{3}\right) \ .  \label{resultb}
\end{equation}%
In the sequel we will drop the tilde notation on the renormalized
wavefunctions.

We could have obtained the same result directly from the ground state
wavefunction (\ref{Suth}) with the coupling $\lambda =1$. In this case the
Sutherland model corresponds to a theory of free fermions on a circle.%
\footnote{%
A connection between Chern-Simons theory and free fermions at finite
temperature is discussed in~\cite{Tierz:2008vh}.} Furthermore, the
configuration employed, with particles equally spaced on the circle,
corresponds to the static classical equilibrium state of the Sutherland
model. This condition is essential to preserve integrability when the model
is discretized to construct spin chain models (see~\cite{Poly} for a
review). It follows that we can also write the free energy of Chern-Simons
theory on $S^3$ as%
\begin{equation}
F_{N,k}\left(S^{3}\right)=\lim_{\lambda \rightarrow \infty }\,\frac{1}{%
\lambda }\, \log \Psi _{0}\left( q_{1},\dots,q_{N};\lambda,k+N\right) \ .
\end{equation}

Recall that the static classical equilibrium configuration of the Sutherland
model, together with an additional spin interaction term in the Hamiltonian,
is the one that leads to the Haldane-Shastry spin chain model~\cite%
{Haldane,Shastry}.\footnote{%
More generally, one can include internal $SU(N)$ colour degrees of freedom
for the particles.} Hence the part without the internal spin degrees of
freedom naturally induces the Chern-Simons partition function. However, the
translationally invariant spin wavefunction~\cite[eq.~(5)]{SC}%
\begin{equation}
\check\Psi _{0}(s_{1},\dots ,s_{N};\alpha ,N)=\delta
_{\sum_{n}\,s_{n}\,,\,0}~{\,\mathrm{e}}\,^{{\,\mathrm{i}\,}\frac{\pi }{2}%
\,\sum_{i\geq 0}\,(s_{2i+1}-1)}\,\prod\limits_{1\leq m<n\leq N}\,\left( \sin 
\frac{\pi \left( n-m\right) }{N}\right) ^{\alpha \,s_{n}\,s_{m}}\ ,
\end{equation}%
with $s_{n}=\pm \,1$ local spins and $\alpha $ a positive real number, can
be expressed in terms of hard-core boson variables. When $N$ is even, one
can identify the spin up (resp.~down) states with empty (resp.~occupied)
hard-core boson states to write~\cite[eq.~(6)]{SC}%
\begin{equation}
\check\Psi^{\mathrm{bos}}_{0}(n_{1},\dots ,n_{\frac{N}{2}};\alpha ,N)={\,%
\mathrm{e}}\,^{{\,\mathrm{i}\,}\pi \,\sum_{i}\,n_{i}}\,\prod\limits_{\overset%
{\scriptstyle1\leq i,j\leq \frac{N}{2}}{\scriptstyle n_{i}>n_{j}}}\,\left(
\sin \frac{\pi \left( n_{i}-n_{j}\right) }{N}\right) ^{4\alpha }\ ,
\end{equation}%
where $n_{i}=1,\dots ,N$ denote the positions of $\frac{N}{2}$ hard-core
bosons in the periodic chain. According to our analysis above, this also
yields the Chern-Simons partition function on $S^{3}$.

\subsection{Framing dependence\label{Framing}}

The many-body quantum system with ground state wavefunction%
\begin{equation}
\Psi _{0}^{\mathrm{CS}}(q_{1},\dots ,q_{N};g_{s},L)=\sqrt{\alpha _{N}}%
~\prod_{i=1}^{N}\,\mathrm{e}^{-{q_{i}^{2}}/{2g_{s}}}~\prod_{i<j}\,\sinh 
\frac{q_{i}-q_{j}}{2L}  \label{CS}
\end{equation}%
is the one that appears in Chern-Simons theory~\cite{Tierz:2008vh}, with the
normalization constant $\alpha _{N}$ giving the Chern-Simons partition
function when $L=1$~\cite{Tierz}. It is also possible to obtain from (\ref%
{CS}) the Chern-Simons partition function $Z_{N,k}(S^{3})$ in the same way
as above. This was already carried out in~\cite{dH} in the context of
Brownian motion (see also~\cite[Appendix~B]{Kapustin:2009kz}).

The relevant density considered in~\cite{dH} is given by%
\begin{equation}
\Psi _{0}^{\mathrm{Brown}}(q_{1},\dots ,q_{N};t)=\prod_{i=1}^{N}\,\mathrm{e}%
^{-{q_{i}^{2}/t}}~\prod_{i<j}\,\sinh \frac{q_{i}-q_{j}}{2} \ ,
\label{CS-Brown}
\end{equation}%
with $t=-1/g_{s}$. The Gaussian factor in $\left( \ref{CS-Brown}\right) $
only contributes to a part of a framing factor~\cite{cs,Jeffrey:1992tk} when 
$q_{1},\dots ,q_{N}$ are taken to be the components of the Weyl vector $\rho$
of $U(N),$ which is given by 
\begin{equation}
\rho =\sum_{i=1}^{N}\, \left( \frac{N+1}{2}-i\right) \, e_{i}
\end{equation}
with $e_{i}$ the standard basis of unit vectors in $\mathbb{R}^{N}$. This is
simply the equilibrium configuration of the Sutherland model that we
considered above. The contribution of the Gaussian factor is then the phase~%
\cite{dH}%
\begin{equation}
\mathrm{e}^{-\left\vert \rho \right\vert ^{2}/{t}}=\exp \left( \frac{{\,%
\mathrm{i}\,}\pi \,N\left( N^{2}-1\right) }{12\left( k+N\right) }\right) \ .
\label{framing}
\end{equation}%
Thus the model that leads directly to the Chern-Simons partition function
with the canonical framing on $S^{3}$ is the Sutherland model, while the
model with ground state wavefunction $\left( \ref{CS-Brown}\right) $ leads
to the Chern-Simons partition function in the matrix model framing~\cite{dH}.

Generally, the contribution of a framing $\Pi $ on a three-manifold $M$
(here $M=S^3$), i.e. a choice of trivialization of the bundle $TM\oplus TM$,
to the partition function for gauge group $G$ (here $G=U(N)$) is
parametrized by an integer $s\in\mathbb{Z}$ and given by~\cite%
{Jeffrey:1992tk}%
\begin{equation}
\delta( M,\Pi) =\exp \Big( \frac{2\pi {\,\mathrm{i}\,} s}{24}\, c\Big) =\exp %
\Big( \frac{2\pi {\,\mathrm{i}\,} s}{24}\, \frac{k\,\dim G}{k+h}\Big) =\exp %
\Big( \frac{\pi {\,\mathrm{i}\,} s\left\vert \rho \right\vert ^{2}k}{h\,
(k+h)}\Big) \ ,  \label{framJeffrey}
\end{equation}%
where $h$ is the dual Coxeter number of $G$ (here $h=N$). In the second
equality here we have used the explicit expression for the central charge $c$
of the associated Wess-Zumino-Witten conformal field theory based on the
affine extension of $G$. In the third equality we have used the
Freudenthal-de~Vries formula to relate the dimension of the Lie group $G$ to
the length of its Weyl vector $\rho$ (see~\cite{Jeffrey:1992tk} for
details). Thus if we write the framing factor (\ref{framJeffrey}) in the form%
\begin{equation}
\delta( M,\Pi) =\exp \Big( \frac{\pi {\,\mathrm{i}\,} s\left\vert \rho
\right\vert ^{2}}{h}\Big)\, \exp \Big( -\frac{\pi {\,\mathrm{i}\,}
s\left\vert \rho \right\vert ^{2}}{k+h}\Big) \ ,  \label{twofactors}
\end{equation}%
we see that the contribution (\ref{framing}) is given by the second factor
in (\ref{twofactors}).

A phase contribution that does not include the level $k$ of the Chern-Simons
gauge theory, as in the first factor of (\ref{twofactors}), appears when one
goes from the hyperbolic to the trigonometric case and conversely. Indeed,
one can also consider the hyperbolic Sutherland model with ground state
wavefunction%
\begin{equation}
\Psi _{0}^{\mathrm{hyp}}(q_{1},\dots ,q_{N};\lambda
,L)=\prod\limits_{i<j}\,\left( \sinh \frac{\pi \left( q_{i}-q_{j}\right) }{L}%
\right) ^{\lambda }\ ,
\end{equation}%
which coincides with that of the Chern-Simons model $\left( \ref{CS}\right) $
at $\lambda =1$ and without the Gaussian factors in the wavefunction. Then
one obtains a relation analogous to $\left( \ref{resultb}\right) ,$ with an
additional phase factor due to the appearance of hyperbolic sine functions
instead of trigonometric ones (this factor also arises when one uses $\left( %
\ref{CS-Brown}\right) $). Repeating the procedure in (\ref{evaluation})
using the relation $\sinh( {\,\mathrm{i}\,} x) ={\,\mathrm{i}\,}\sin x$ we
see that we now have to take $L=-\mathrm{i}\, (k+N)$ and that the additional
phase factor, including the corresponding renormalization (\ref{waveren}),
is $\mathrm{\exp }\left( \frac{1}{4}{\,\mathrm{i}\,}\pi \,N^{2}\right) $,
which also appears in the random matrix theory computation of $\alpha _{N}$~%
\cite{Tierz}.

\subsection{Extension to generic root systems}

The results we have described are not restricted to unitary gauge groups,
and can be directly extended to orthogonal and symplectic gauge groups using
generalizations of the Sutherland model appropriate to other semi-simple Lie
algebras~\cite{Perelo}. The generalized Calogero-Sutherland model is defined
by the quantum Hamiltonian operator~\cite{Perelo}%
\begin{equation}
H=-\sum_{i=1}^{N}\,\frac{\partial ^{2}}{\partial q_{i}^{2}}+\sum_{\alpha \in
\Delta }\, \frac{g_{\left\vert \alpha \right\vert }}{\sin ^{2}\frac12\,(q,
\alpha) } \ ,
\end{equation}%
where $\Delta $ is the root system of a semi-simple Lie algebra in an $N$%
-dimensional real vector space $V$ with inner product $(-,-)$, $%
q=(q_1,\dots,q_N)\in V$, and $g_{\left\vert \alpha \right\vert
}=\lambda_{|\alpha|}\,(\lambda_{|\alpha|}-1)\,\pi^2/L^2$ are coupling
constants which depend only on the lengths $|\alpha|$ of the root vectors $%
\alpha\in\Delta$ (i.e. on the orbits of the Weyl group). The usual
Sutherland model (\ref{SuthHam}) corresponds to the $A_{N-1}$ root system.

The $BC_{N}$ root system contains three distinct Weyl group orbits and hence
three coupling constants. The Hamiltonian is%
\begin{align}
H=& -\sum_{i=1}^{N}\, \frac{\partial ^{2}}{\partial q_{i}^{2}}+\frac{%
2\lambda \left( \lambda -1\right) \pi ^{2}}{L^{2}}\, \sum_{i<j}\, \left( \sin%
\frac{\pi \left( q_{i}-q_{j}\right) }{L}\right) ^{-2}  \notag \\
& +\, \frac{2\lambda \left( \lambda -1\right) \pi ^{2}}{L^{2}}\,
\sum_{i<j}\, \left( \sin\frac{\pi \left( q_{i}+q_{j}\right) }{L}\right)^{-2}
\notag \\
& +\, \frac{\lambda _{2}\left( \lambda _{2}-1\right) \pi ^{2}}{L^{2}}\,
\sum_{i=1}^{N}\, \left(\sin \frac{\pi\, q_{i}}{2L}\right)^{-2} + \frac{%
\lambda _{1}\left( \lambda _{1}-1\right) \pi ^{2}}{L^{2}}\, \sum_{i=1}^{N}\,
\left(\sin\frac{\pi\, q_{i}}{L}\right)^{-2} \ ,
\end{align}%
with corresponding ground state wavefunction%
\begin{align}
\Psi _{0}( q_{1},\dots,q_{N};\lambda_1,\lambda_2,\lambda,L) =&
\prod\limits_{i=1}^{N}\, \sin ^{\lambda _{1}}\left( \frac{\pi\, q_{i}}{L}%
\right)\, \sin ^{\lambda _{2}}\left( \frac{\pi \, q_{i}}{2L}\right)
\label{CN} \\
& \times\, \prod\limits_{i<j}\, \left( \sin \frac{\pi \left(
q_{i}-q_{j}\right) }{2L}\right) ^{\lambda }\, \left( \sin \frac{\pi \left(
q_{i}+q_{j}\right) }{2L}\right) ^{\lambda } \ .  \notag
\end{align}%
The case $\lambda _{1}=0$ corresponds to the $B_{N}$ root lattice, $\lambda
_{2}=0$ to the $C_{N}$ root lattice, and $\lambda _{1}=\lambda _{2}=0$ to
the $D_{N}$ root lattice. This generalized Sutherland model is related, as
in the case of the $A_{N-1}$ system, to the partition function of
Chern-Simons gauge theory on $S^{3}$ with orthogonal and symplectic gauge
groups. The generic framing dependence is described by~(\ref{twofactors}).

For illustration, let us consider explicitly the Lie group $Sp(2N).$ The
configuration is again fixed and equispaced as%
\begin{equation}
\left( q_{1},\dots ,q_{N}\right) =\left( N,N-1,N-2,\dots ,1\right) \ ,
\end{equation}%
and this choice corresponds to the Weyl vector $\rho$ of the Weyl chamber of 
$C_{N}$. Note that there is no essential difference with the $A_{N-1}$
system treated earlier, where the equispaced configuration was taken to be $%
q_{i}=c-i$ with arbitrary but fixed $c\in \mathbb{R}$. Evaluating $\left( %
\ref{CN}\right) $ with $\lambda =\lambda _{1}=\frac{1}{2}$ and $\lambda
_{2}=0$ yields the probability density~\cite{dH}%
\begin{equation}
\big(\Psi _{0}(N,N-1,\dots ,1;\mbox{$\frac12,0,\frac12$},L)\big)%
^{2}=\prod\limits_{j=1}^{2N+1}\,\left( \sin \frac{j}{2L}\right) ^{f(j)}\ ,
\end{equation}%
where 
\begin{equation}
f(j)=\left\{ 
\begin{array}{cc}
N-\mbox{$\frac j2-\frac12$} & j\text{ odd}\leq N \\ 
N-\mbox{$\frac j2$} & j\text{ even}<N \\ 
N-\mbox{$\frac j2+\frac12$} & j\text{ odd }>N \\ 
N-\mbox{$\frac j2$}+1 & j\text{ even }>N\ .%
\end{array}%
\right.
\end{equation}%
With the identification (\ref{LkN}), this is the partition function of
Chern-Simons theory on $S^{3}$ with gauge group $Sp(2N)$~\cite{Sinha:2000ap}.

\subsection{Derivation of Wilson line observables}

The Sutherland model also delivers other quantum topological invariants,
besides partition functions, in a similar way. For example, quantum
dimensions~\cite{Fuchs} can be obtained in this way as well. In Chern-Simons
gauge theory, the Wilson line invariants associated to the unknot give rise
to the quantum dimensions which reduce, in the semiclassical limit $%
k\rightarrow \infty $ ($g_s\to0$), to the dimensions of representations of
the gauge group. For example, the irreducible representations $R$ of the
gauge group $SU(N)$ can be parametrized by the lengths of the rows of Young
tableaux $\mu_{i},$ $i=1,\dots,N$, with $\mu_{1}\geq \mu_{2}\geq \cdots\geq
\mu_N$. Using again $\left( \ref{waveren}\right) $ and $\left( \ref%
{evaluation}\right) $, but now with these weights as the positions $%
q_i=\mu_i $ of the particles, i.e. an inhomogeneous configuration, one
obtains%
\begin{equation}
\dim _{q}R=W_{R}( \mathrm{unknot}) =\frac{1}{L^{N/2}}\, \prod\limits_{i<j}\,
\sin \frac{\pi \left( \mu_{i}-\mu_{j}\right) }{L}
\end{equation}%
again with the identification (\ref{LkN}). In this way knot and link
invariants in Chern-Simons gauge theory correspond to non-equilibrium
configurations of particle positions in the Sutherland model. It would be
interesting to study excited states of the Sutherland model in its classical
equilibrium configuration to see if they give rise to quantum dimensions or
other observables of Chern-Simons theory.

\subsection{Hamiltonian analysis}

Recall that the hyperbolic Sutherland model and the Chern-Simons model
differ in a Gaussian factor at the level of the ground state wavefunction (%
\ref{CS}). In the relationship with Chern-Simons theory, this is only a
(framing) phase contribution to the partition function. At the level of the
corresponding Hamiltonians, they are related by~\cite{Tierz:2008vh}%
\begin{equation}
H_{\mathrm{CS}}=H_{\mathrm{hyp}}+\frac{1}{g_{s}\,L}\,\sum_{i<j}\left(
q_{i}-q_{j}\right) \coth \left( \frac{q_{i}-q_{j}}{L}\right)
\end{equation}%
where $H_{\mathrm{hyp}}$ is the Hamiltonian of the hyperbolic Sutherland
model. The reason for the appearance of a two-body term at the Hamiltonian
level to explain a one-body factor in the ground state wavefunction is
explained in~\cite{Tierz:2008vh} (together with some other properties of the
model).

At large distances, the two-body potential is given by a one-dimensional
Coulomb potential $V(x-x^{\prime }\,)=\left\vert x-x^{\prime }\,\right\vert $%
. Thus it has strongly confining properties, and while both the Sutherland
model and the Chern-Simons model are connected to Chern-Simons gauge theory,
the addition of such a potential leads to a very different correspondence.
For example, in~\cite{FradMoreno} the equivalent relationship between the
Sutherland model and the Thirring model is presented, and a generalized
Sutherland model with a two-body potential is also studied. It is shown
there that in order for the correspondence between the Sutherland and
Luttinger-Thirring models to hold, the generic two-body potential $V(x)$
cannot decay slower than $\left\vert x\right\vert ^{-2}$ for large~$x$.%
\footnote{%
This ensures that the Fourier transform of the potential is not of the form $%
\tilde{V}(k)=A/k^{\sigma }$ with $\sigma >1$ (see~\cite{FradMoreno} for
details).} As explained in~\cite{FradMoreno}, the one-dimensional Coulomb
potential has Fourier transform of the form $\tilde{V}(k)=A/k^{2}$ and
the physics is that of one-dimensional quantum electrodynamics, rather than
the Luttinger-Thirring behaviour of the Sutherland model. Indeed, a linear
confining potential (such as the one-dimensional Coulomb potential) is one
of the main ingredients of a massive Schwinger model. It would be
interesting to understand further the implications of the fact that the
two-body potential of the Chern-Simons model is, at large separations, a
one-dimensional Coulomb potential. The exactly solvable Chern-Simons model
is also related to other systems in condensed matter physics~\cite{T2}, such
as Laughlin and multilayer wavefunctions on thin cylinders, and also through
a connection between the Sutherland model and Luttinger liquids with
boundaries.


\end{document}